\documentclass[aps,prl,twocolumn,showpacs,preprintnumbers,amsmath,amssymb]{revtex4-1}

\usepackage{graphicx}
\usepackage{color}
\usepackage{upgreek}
\usepackage{float}
\usepackage{amsmath}

\def\per{$\rm [\bar{1} \bar{1} 2]\,$}

\begin{document}
\title{Two-dimensional crossover and strong coupling of plasmon excitations in arrays of one-dimensional atomic wires}

\author{T. Lichtenstein$^1$}  \author{J. Aulbach$^3$} \author{J. Sch\"afer$^3$} \author{R. Claessen$^3$}
\author{C. Tegenkamp$^{1,2}$}
\author{H. Pfn\"ur$^{1,2}$} \email{pfnuer@fkp.uni-hannover.de}

\affiliation{$^1$ Institut f\"ur Festk\"orperphysik, Leibniz Universit\"at
Hannover, Appelstra\ss e 2, 30167 Hannover, Germany}

\affiliation{$^2$ Laboratorium f\"ur Nano- und Quantenengineering (LNQE),
Leibniz Universit\"at Hannover, Schneiderberg 39, 30167 Hannover, Germany}

\affiliation{$^3$ Physikalisches Institut and RCCM, Universit\"at W\"urzburg, Am Hubland, W\"urzburg, Germany}

\date{\today}

\begin{abstract}
The collective electronic excitations of arrays of Au chains on regularly stepped Si(553) and Si(775) surfaces were
studied using electron loss spectroscopy with simultaneous high energy and momentum resolution (ELS-LEED) in 
combination with low energy electron diffraction (SPA-LEED) and tunneling microscopy. 
Both surfaces contain a double chain of gold atoms per terrace. 
Although one-dimensional metallicity and plasmon dispersion is observed only along the wires, 
two-dimensional effects are important, since plasmon dispersion explicitly
depends both on the structural motif of the wires and the terrace width.
The electron density on each terrace turns out to be modulated, as seen by tunneling spectroscopy (STS).
The effective wire width of  7.5\,\AA\ for Si(553)-Au -- 10.2\,\AA\ for Si(775)-Au -- ,
determined from plasmon dispersion is in good agreement with STS data. 
Clear evidence for coupling between wires is seen beyond nearest neighbor coupling. 
\end{abstract}
\pacs{73.50.Jt, 68.43.-h, 75.23.Lp}
\maketitle
%************************************************************************
% Introduction
%\section{Introduction}
Plasmons, i.e. the collective excitations of electrons, play an important role, e.g. in 
sensor technology \cite{Schwarz14}, improvement of quantum efficiency in photovoltaic devices \cite{Atwater10},
and  even in cancer research \cite{Oroczo14}.  Recently,  collective excitations of low-dimensional electron gases, 
called sheet plasmons, came into focus of research \cite{Nagao10,Vattuone13}. 
The wavelength of these sheet plasmons is typically three orders of magnitude  shorter 
compared to photons of the same frequency.  Thus THz-plasmonics on the scale of a few nanometers becomes feasible. 

One-dimensional (1D) metallic wires and their plasmonic excitations would be ideal for directed energy transport
on the nanoscale, since quasi-linear dispersion is predicted, at least in the long wavelength limit \cite{Sarma85}, 
for these 1D plasmons. Such dispersions have indeed been found for regular arrays of atomic wires on insulating 
substrates \cite{Nagao07,Rugeramigabo10,Krieg13}. 
Under certain conditions of coupling and mutual screening  
linearized plasmon dispersions, so-called acoustic surface plasmons, are obtained in 2D systems \cite{Diaconescu07, Vattuone13}. 
Moreover, confinement effects in these metallic subunits on the surface lead to formation of intersubband 
excitations \cite{Rugeramigabo10,Krieg13,Smerieri14}. On the other hand, several fundamental aspects of these low-D plasmons 
such as many-body effects, electronic correlations and Coulomb screening \cite{Lindhard54,Stern67,Singwi68} 
are still rather unexplored and lead to an unsatisfactory description of experimental results 
\cite{Inaoka07,Nagao07,Rugeramigabo10}.

Growth of various metals in the submonolayer regime on semiconducting surfaces  provide a superb approach 
for addressing such fundamental aspects for 1D and 2D systems. The adsorbate induced band structure is generally electronically 
decoupled from the bulk bands of the host material. Au chains on regularly stepped Si(111) surfaces 
at various tilt angles towards the \per direction are no exception and are particularly interesting systems because of intrinsic surface magnetism 
\cite{Erwin10, Aulbach13}. 
Depending on coverage and vicinality, the widths of the Au-chains and their interwire 
spacing can be tuned, while their electronic band structures are still very similar. 
0.2\,ML of Au on Si(557), e.g., result in growth of single atom Au-chains and a row of Si-adatoms 
on each mini-terrace with an interwire spacing of 19.2\,\AA\ \cite{Crain04,Nagao06}. In contrast, Si(553) and Si(775)-Au 
host double Au chains in the center of the terrace \cite{Aulbach15,Krawiec10}. 
The interwire spacing is 14.8\,\AA\ for Si(553)  and 21.3\,\AA\ for Si(775) \cite{Crain04,Barke09,Erwin10}. For double Au chains, 
nominal coverages of 0.48\,ML on Si(553) and of 0.32\,ML on Si(775) result.  
Common to all these structures is a graphitic Si-honeycomb chain located at the step edges \cite{Crain04,Barke09,Erwin10}.
%As known from angular resolved photoemission (ARPES) \cite{Crain04,Barke09,Erwin10}, 
%the $\rm sp^2$-hybridized  Si-honeycomb chain  located 
%at the edges of the monatomic steps is common to all these structures. 
Each of these system is characterized by metallic bands that are well known from angular resolved photoemission (ARPES) measurements \cite{Crain04}. 
They only disperse along 
the chain direction $k_\parallel$, and have their minima at the zone boundary. Thus, also the (equilibrium) electron density available for plasmonic excitations is well known.

In fact, these systems form locally the narrowest possible 1D objects that can be realized, namely chains that are 
one or two atoms wide. Therefore, we address here the question of local confinement both for the ground state
close to the Fermi level and for the collective excited plasmon state with emphasis on the plasmons. 
We compare the collective excitations in Si(553)-Au and Si(775)-Au, since the same structural motif of the 
double gold chain is present in both systems, and make reference to the Si(557)-Au system with only a single gold chain. 
Although purely 1D dispersion along the chain direction is found, the lateral extension of the 
charge distribution turns out to explicitly influence the slope of the measured plasmon dispersion curves. 
In other words, this crossover into the second dimension is crucial for the quantitative interpretation of a 1D phenomenon, but
is not described by existing theories. 
In contrast, the plasmonic coupling between the wires in the ordered arrays, which is another aspect of dimensional crossover, 
can be described  quantitatively by existing mesoscopic theories \cite{Li90,DasSarma96}. 

All experiments were performed in two different ultra-high vacuum chambers operating at a base pressure of $5 \times 10^{-11}\,\text{mbar}$. 
One system hosts a high resolution spot profile analysis low energy electron diffractometer 
(SPA-LEED) to investigate and control the sample quality, and a combination of an electron energy loss 
spectrometer with a LEED diffractometer providing high resolution both in energy and momentum \cite{Claus92} in order to determine
plasmon dispersion relations. The overall sample quality was checked by a SPA-LEED.
The vicinal Si-substrates ($\rho \approx 0.01\,\Omega\text{cm}$, n-type) were annealed at 
$1250\,^\circ\text{C}$ for a few seconds 
followed by rapid cool down. 
The appropriate coverages of 0.48\,ML for Si(553)-Au and 0.32\,ML for Si(775)-Au were evaporated from a gold pearl on a tungsten 
filament by direct current heating, or from a crucible at a substrate temperature of $630\,^\circ\text{C}$ . 
The coverage has been controlled and calibrated by quartz microbalances placed at the position of the samples 
\cite{Sauerbrey59}. After Au-deposition and cooling room temperature the samples were quickly annealed to 
$930\,^\circ\text{C}$ for $< 1$\,s,  followed by instantaneous cooling to room temperature.
The loss measurements were carried out directly after this post-annealing step in order to avoid 
any influence of residual gas on surface and electronic structure. 
The scanning tunneling microscopy (STM) measurements were performed at 77\,K in the second chamber using a low temperature STM manufactured 
by Omicron. The overall sample quality here was checked with an optical LEED.

%************************************************************************
% Results and discussion
%\section{Results and discussion}
%\subsection{Atomic structure: LEED and STM}
%
\begin{figure}
\begin{center}%
\includegraphics[width=0.8\columnwidth]{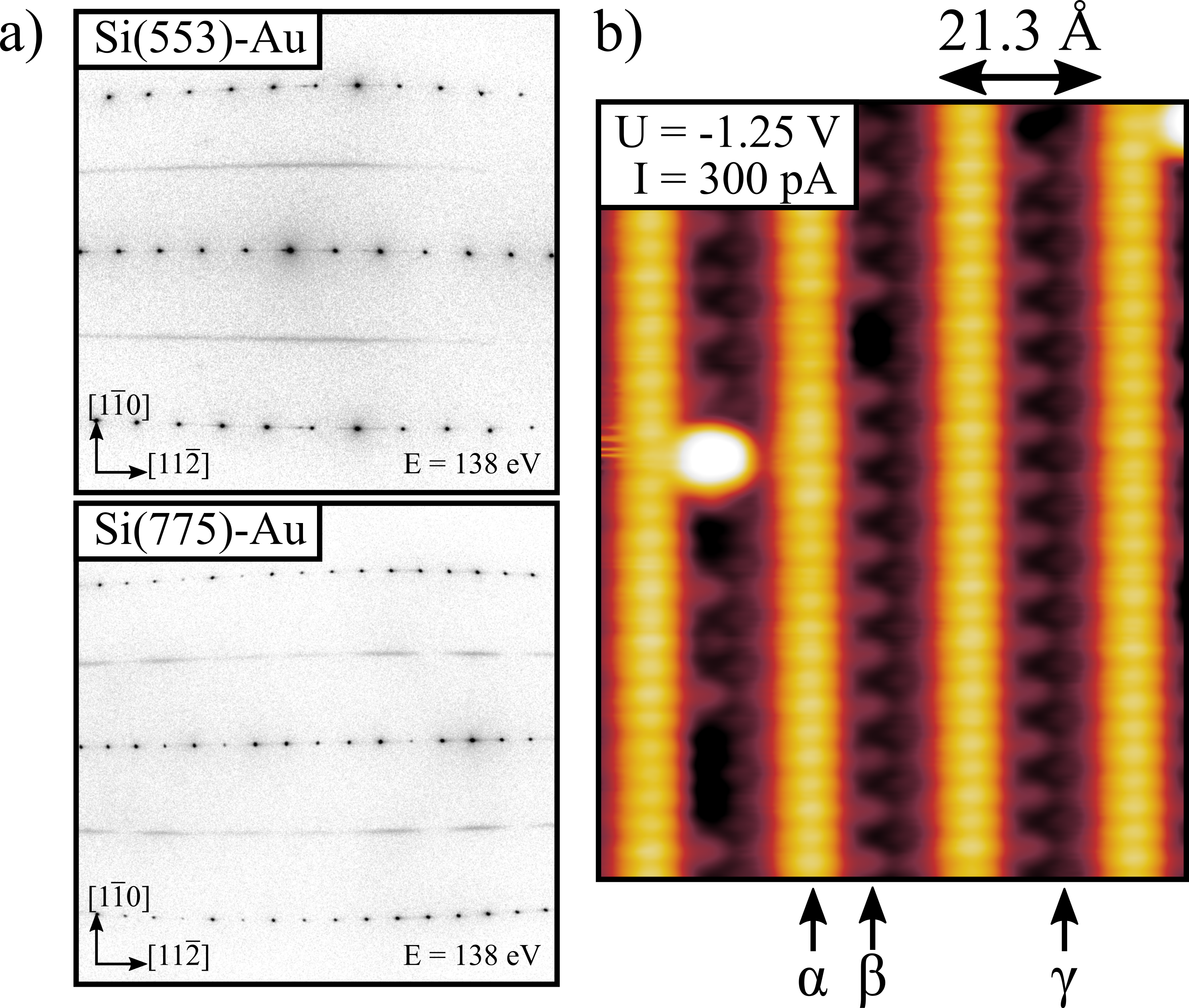}
		\caption{\label{FIG1}
		(color online) a) LEED patterns of the chain structures of Si(553)-Au and Si(775)-Au. 
		b) STM image of the Si(775)-Au 
		chain structure	(tunneling conditions U= -1.25 V, I = 300 pA): $\upalpha, \upbeta, \upgamma$ mark the Si terrace 
		edge, a row of Si adatoms accompanied by Si restatoms, and a Au double chain with $\times 2$ period doubling, respectively \cite{Aulbach15}.
		STM experiments were performed at 77\,K, LEED at 300 K. Bright protrusion could be identified as an adatom defect \cite{Aulbach15}.}
	\end{center}
\end{figure}

Here we demonstrate the close structural similarities between the Si(553)-Au and Si(775)-Au systems which complement and fit to 
the very similar electronic structures mentioned above. The LEED patterns right after preparation are shown in fig.\,\ref{FIG1}a). 
These patterns are characteristic for regularly stepped (553) and (775) surfaces. They consist of (111)-oriented 
terraces and steps of double atomic height ($d=3.14$\,\AA). From the spot splitting of 22\,\% SBZ (surface Brillouin zone) 
for Si(553)-Au and 15.6\,\% SBZ for Si(775) we derive average terrace widths of $14.8\,\text{\AA}$ 
for the Si(553) and of $21.3\,\text{\AA}$ for the Si(775) surface. Thus, the adsorption of 0.48\,ML of Au on Si(553) -- 
0.32\,ML on Si(775) -- leaves the periodic array of double steps with sharp spots and $\times 2$ streaks unchanged, 
indicating high quality 1D order. This is in agreement with STM, exemplarily shown for the (775)-system in (fig.\,\ref{FIG1}c)).
A narrow distribution of terrace widths can be concluded from $(k_{||},k_{\perp})$-plots in LEED (not shown) and from STM.

High resolution STM, shown in fig.\,\ref{FIG1}b) for the Si(775)-Au system, reveals details of the atomic arrangement (for an extended set 
of STM data on Si(553)-Au, see ref.\,\cite{Aulbach13}). 
Each terrace hosts three structural motifs. Their origin could be disentangled by a detailed analysis via STM and density functional 
theory, which will be published elsewhere \cite{Aulbach15}. However, we will list our results here: 
The motif denoted by $\alpha$ could be ascribed to the Si honeycomb chain at the
step edge. Chain $\beta$ is formed by a Si adatom row accompanied by Si restatoms (i.e. unpassivated Si atoms).  
Motif $\gamma$, most important for this work, could be identified as a double Au-strand \cite{Aulbach15}, 
similar to what was found in Si(553)-Au \cite{Erwin10,Aulbach13,Krawiec10}.

Most importantly, the chain structures show a $2a=7.7\,\text{\AA}$ 
periodicity along the wire direction. Correspondingly, LEED reveals modulated $\times 2$
diffraction streaks. The streaks along the \per direction are indicative for  only 
short-range  correlation between the double periodicity along the chains on the different terraces.
For Si(553)-Au, only the Au chains show the $\times 2$ periodicity. 
Therefore, the modulation and intensity of the $\times 2$ diffractions streaks there are weaker.

%In fig.\,\ref{FIG1}b) two types of structural defects are also visible: 
%adatoms that appear as bright protrusions, and dark holes between the bright rows 
%that are identified as defects within the gold strands.

%\subsection{1D excitations in Au wires}

\begin{figure}[tb]
	\begin{center}
\includegraphics[width=.9\columnwidth]{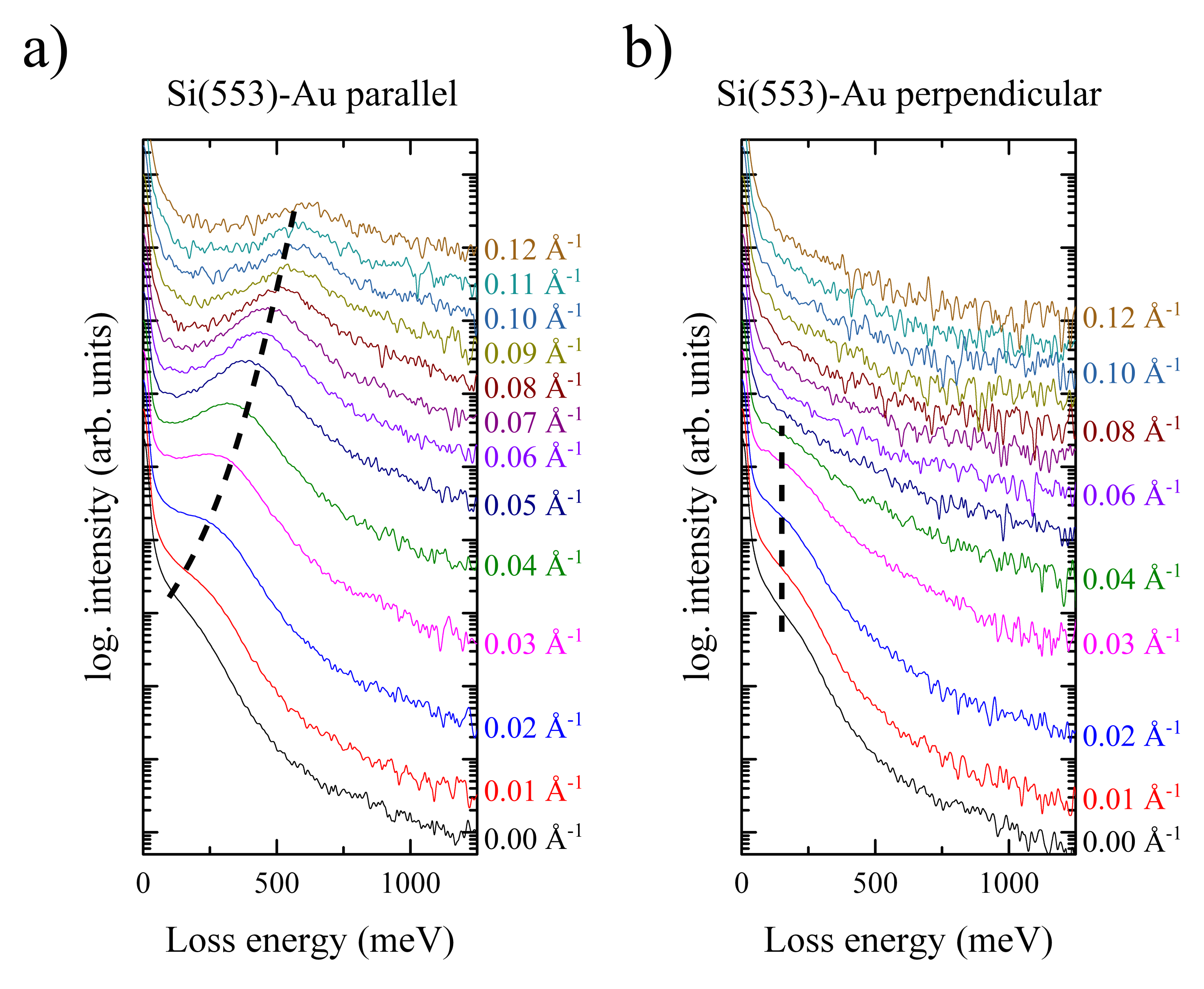}
		\caption{\label{FIG2}
		(color online) Electron energy loss spectra (primary energy 20\,eV) of Au-chains grown on Si(553) 
		as a function of $k$ parallel and normal to the wires, as indicated.}
	\end{center}
\end{figure}

On these well ordered arrays of 1D atomic chains, angle resolved electron energy loss spectroscopy measurements 
were performed. Fig.\,\ref{FIG2}a) shows sequences of loss spectra on semi-log scale as a function of 
increasing $k_\parallel$ for Si(553)-Au. A similar plot for the orthogonal direction is shown in 
fig.\,\ref{FIG2}b). Corresponding spectra for Si(775)-Au are shown in the supplement. 
Close to $k_\parallel = 0$ the spectra are structureless, apart from a small 
non dispersing feature that dies out quickly with increasing $k_\parallel$.
The exponentially decaying loss intensity as a function of loss energy elastic peak, known as Drude tail, is the typical signature  
of the continuum of low-energy excitations in metallic systems \cite{Krieg15}. 
It provides evidence that both systems are metallic, in agreement with 
findings from ARPES for the Si(553)-Au \cite{Crain04}, STM \cite{Aulbach13}, and theory \cite{Erwin10}.
However, it is at variance with the ARPES data for the Si(775)-Au \cite{Crain04} for reasons still to be explored.

In the direction along the wires, clear loss features are observed, which shift to 
higher loss energies with increasing scattering angles, i.e. with  increasing $k_\parallel$. 
In the $k_\bot$ direction, however, i.e. in the direction across the wires, no dispersing mode is seen (see fig.\ref{FIG2}b)).
These findings are indicative for the existence of 1D plasmons.

%Commonly, the short lifetimes of these excitations result in comparably broad loss features and 
%necessitate an accurate fitting procedure of the energy loss spectra in order to determine reliably 
%the plasmon dispersion from these sequences. 
Each  spectrum was accurately fitted by parametrizing the elastically scattered peak, the actual 
plasmon loss and the Drude background and by applying the same fitting routine to all spectra. 
Details about the fitting procedure can be found in the supplement and are also elaborated in the appendix of ref.\,\cite{Krieg15}. 
%
%Whereas the intensities of the plasmonic losses in HREELS are typically of the order 
%of 1\% of the elastic peak intensity, we measure here large loss signals signals with 
%an intensity up to 10\% of the elastic peak at off-specular directions. 
%Loss intensities as a function of 
%$k_\parallel$, which reach values up to 10\% of the elastic peak, are shown in fig.\,\ref{FIG2}d. 

We note that  special care has to be taken to eliminate water and hydrogen from the background gas.
Activated by the electron beam, hydrogen causes disorder in the system increasing with time. 
The result is (electronic) break-up of the Au chains, as evident from the appearance 
of losses at finite energy and small $k_\parallel$ whose loss energies increase with 
time. 
%The reduction of the hydrogen partial pressure in the residual 
%gas by one order of magnitude using an additional getter pump reduced this problem correspondingly.   
A similar trend was seen recently also for Ag/Si(557) \cite{Krieg14,Krieg15}.

\begin{figure}
	\begin{center}
		 \includegraphics[width=0.8\columnwidth]{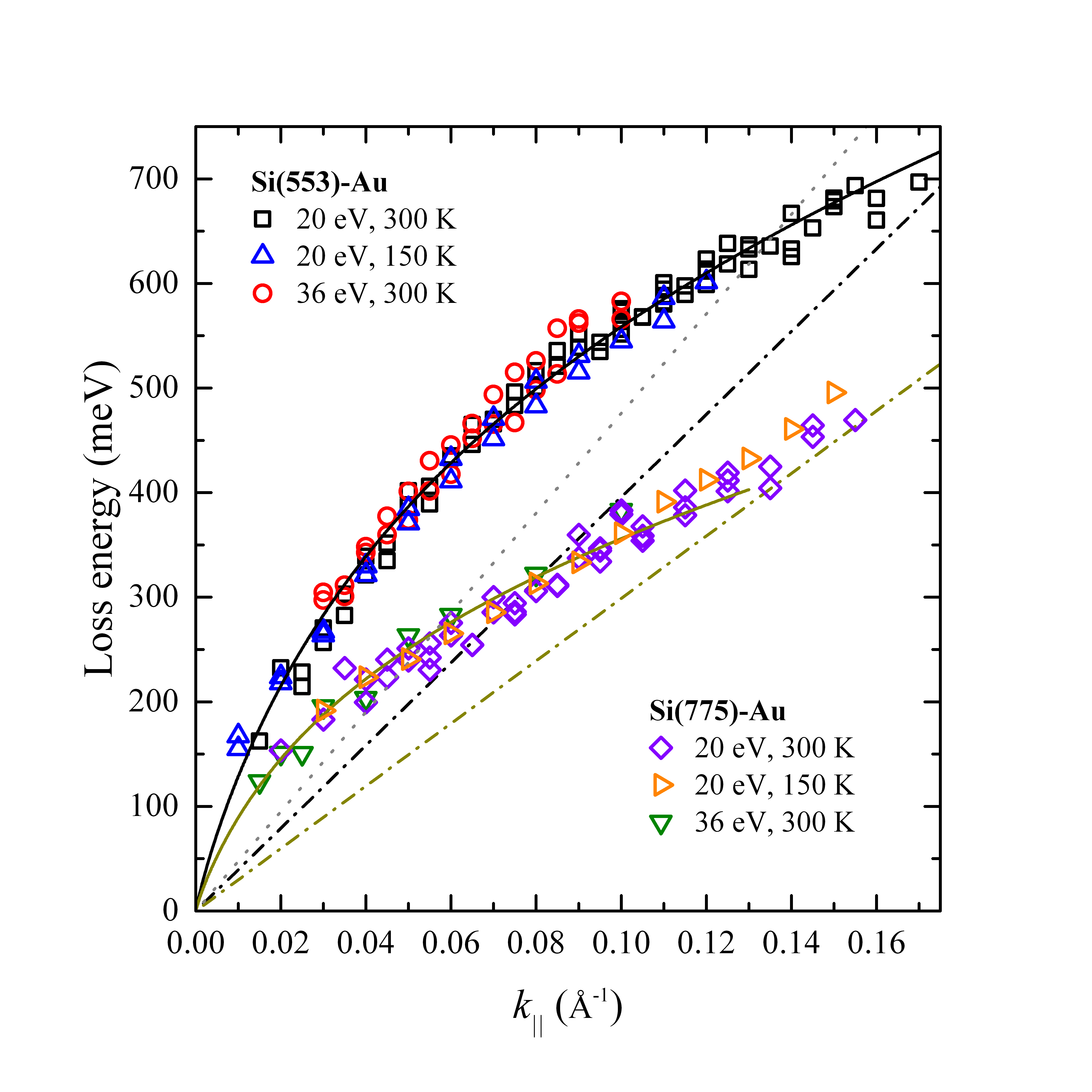}
		\caption{\label{FIG3}
		(color online)  Plasmon dispersion for Au quantum wires grown on Si(553) and on Si(775). Lines are 
		fits according to eq.\ref{eq2}. Dashed-dotted lines: first terms of eq.\,\ref{eq2} for both systems.
		Dotted line: same for Si(557)-Au, fitted data from ref.\,\cite{Nagao06}.
		%The blue curve represents the dispersion found for Si(557)-Au (reproduced from ref.\,\cite{Nagao06}).
		}
	\end{center}
\end{figure}

The  dispersion curves along the wires resulting from the loss maxima of fig.\,\ref{FIG2} are shown in fig.\,\ref{FIG3}. 
Both systems seem to be essentially quasi-1D systems, and should be compared with existing 1D plasmon theory \cite{Sarma85,Li90,Nagao07,Inaoka07}, therefore.
Common to all theoretical approaches is the use of a nearly free electron gas and various approximations for correlations, in the simplest case 
the random phase approximation (RPA). 
With this type of approach, a quantitative fit turned out to be possible with the model of coupled wires, sitting in a periodic array of square potentials at 
distance $d$ \cite{Li90,DasSarma96}. At small $k_\parallel$, the dispersion is given by
\begin{multline}
E = \hbar \sqrt{\frac{4 n e^2}{(1+ \epsilon)\epsilon_0 m^* a^2}} k_\parallel a_0 \times \\
\sqrt{\text{K}_0({\frac{k_\parallel a}{2 \sqrt{2}}) + 2 \sum_{l=1}^L \text{K}_0(k_\parallel ld)\cdot\cos{(k_\bot l d)}}} \label{eq2}
\end{multline}
where the first term of the product contains the electronic and structural properties of a single wire, the second the intrawire 
(first term under square root) and the interwire interaction.
$n$ is the electron density per unit length, $e$ the elementary charge, $m^\star$ the effective mass. 
$\epsilon$ is the dielectric function of Si as partially embedding medium.
$\text{K}_0$ are modified Bessel functions of zeroth order and second kind, $k_\bot$ is the momentum normal to the wires, 
If $a$ (the effective wire width) is set equal to $a_0$ (a constant for normalization), eq.\,\ref{eq2} corresponds to the original formula given in 
refs.\,\cite{Li90,DasSarma96}, which, however, turns out not to describe our findings.
The ratio $a_0/a$ accounts both for differences 
in structural motifs and effective wire widths of a single wire, and is the only free parameter in eq.\,\ref{eq2}.  
In the array of square potentials the first term under the second square root accounts for 
the self-interaction of a single wire, whereas the second term describes the interaction between 
different wires at multiple distances of $d$. 
%%%
%
%The leading term of the dispersion in the limit of (very)  small $k_\parallel a$ ($a$ is the effective wire width) 
%reads for the lowest plasmon mode, $\omega_p$ as a function of $k_\parallel$:
%\begin{equation}
%\omega_p (k_\parallel) = \sqrt{\frac{2ne^2}{m^\star \epsilon \epsilon_0 a^2}} k_\parallel a\cdot ln{(\frac{2\sqrt2}{k_\parallel a})} \label{eq1}
%\end{equation}  

%The dependence on  $a$ thus enters only as a correction for fixed $n$. The same happens in fact for 
%the functional dependence on the terrace separation, $d$ \cite{Li90}.

$n$ and $m^\star$ for the present systems were directly derived from ARPES data \cite{Crain04}, 
i.e. from the occupied band structure of the Au-modified surface states. We also use their surface band notation.
The systems investigated here are characterized by two 
surface bands, an upper $S_1$ and lower $S_2$, crossing the Fermi level \cite{Crain04,Barke09}. As we see a metallic behavior of the Si(775)-Au
system, we assume its bands also to cross the Fermi level for a clean sample at room temperature.
The ratio $n/m^\star$ for both bands is identical within error bars. This means that the plasmons of these bands are degenerate and
cannot be separated experimentally. As a consequence, only a single dispersion curve is expected to be seen for both systems 
with the electron density corresponding to a single band, in agreement with our findings. 

The fits are shown in fig.\,\ref{FIG3} together with the data. 
An explicit dependence of initial slope and shape of dispersion on structural motifs and on terrace widths is found, 
which is not described by existing 1D theories, as we will now demonstrate by concentrating on the first term in eq.\,\ref{eq2}. 

Comparing first Si(553)-Au and Si(775)-Au, which both have the Au double chain, but have different terrace widths,
this first term of eq.\,\ref{eq2} differs by a factor of 1.4 (after correction for small differences of DOS), 
as indicated by the dashed-dotted lines in fig.\ref{FIG3}, i.e., it scales with the inverse of the 
terrace width, $d$ ($21.3\,\text{\AA} / 14.8\,\text{\AA} = 1.43$).

Si(775)-Au and  Si(557)-Au \cite{Nagao06}, on the other hand, have the same terrace widths (within 10\,\%, 19.2 vs. 21.3\,\AA), 
but double and single Au chains per terrace, respectively. As seen from ARPES data \cite{Crain04},  
the 1D electron density of Si(775)-Au in the S2 band is {\it higher} 
by 20\,\% than for Si(557)-Au, whereas the effective masses are virtually 
the same. However, fitting the published data of ref.\,\cite{Nagao06} for Si(557)-Au to eq.\,\ref{eq2}, 
it turns out that its first term is a factor of 1.6 {\it larger} for Si(557)-Au than that of Si(775)-Au. 
Taking the differences in $n$ and the $d$-dependence from above for the two systems, the effective width $a$, as suggested  in eq.\,\ref{eq2}, 
has to be reduced effectively by roughly a factor of 2 for Si(557)-Au compared to Si(775)-Au. 
In other words, not only the periodicity, given by the wire distances $d$, is influencing the dispersion directly, 
but also the {\em internal 2D distribution} of electron density within each wire plays an important role in the plasmon dispersion. 
Since for these narrow structures and the given k$_F$ from ARPES only the lowest subband of a quantum well is occupied, combined excitations
such as intersubband plasmon excitations can be ruled out.

These findings can be summarized by
$$\omega_p \propto \sqrt{\frac{n\cdot a_0^2}{m^\star\cdot a^2}}$$
with $a = \gamma\cdot d$, where $\gamma < 1$ is determined by the 
internal lateral distribution of the electron density in each wire. 
{\em This means that even in case of purely 1D plasmonic dispersion, there is a crossover to 2D, since 
the width of a wire on the atomic scale and the internal electronic distribution within the very wire
enter directly the slope of plasmonic dispersion, which cannot be treated as a correction to 1D properties.} 
  
With respect to coupling between the wires, we obtained the best fit when the sum in eq.\,\ref{eq2} under the 
square root is truncated after the second term ($L=2$), an indication of a finite range of interaction. 
This result may not be quantative, since the model of eq.\,\ref{eq2} neglects damping and dephasing between wires. 
On the other hand, this analysis clearly demonstrates  that the array of 1D plasmons is coupled.

\begin{figure}[tb]
	\begin{center}
		\includegraphics[width=0.7\columnwidth]{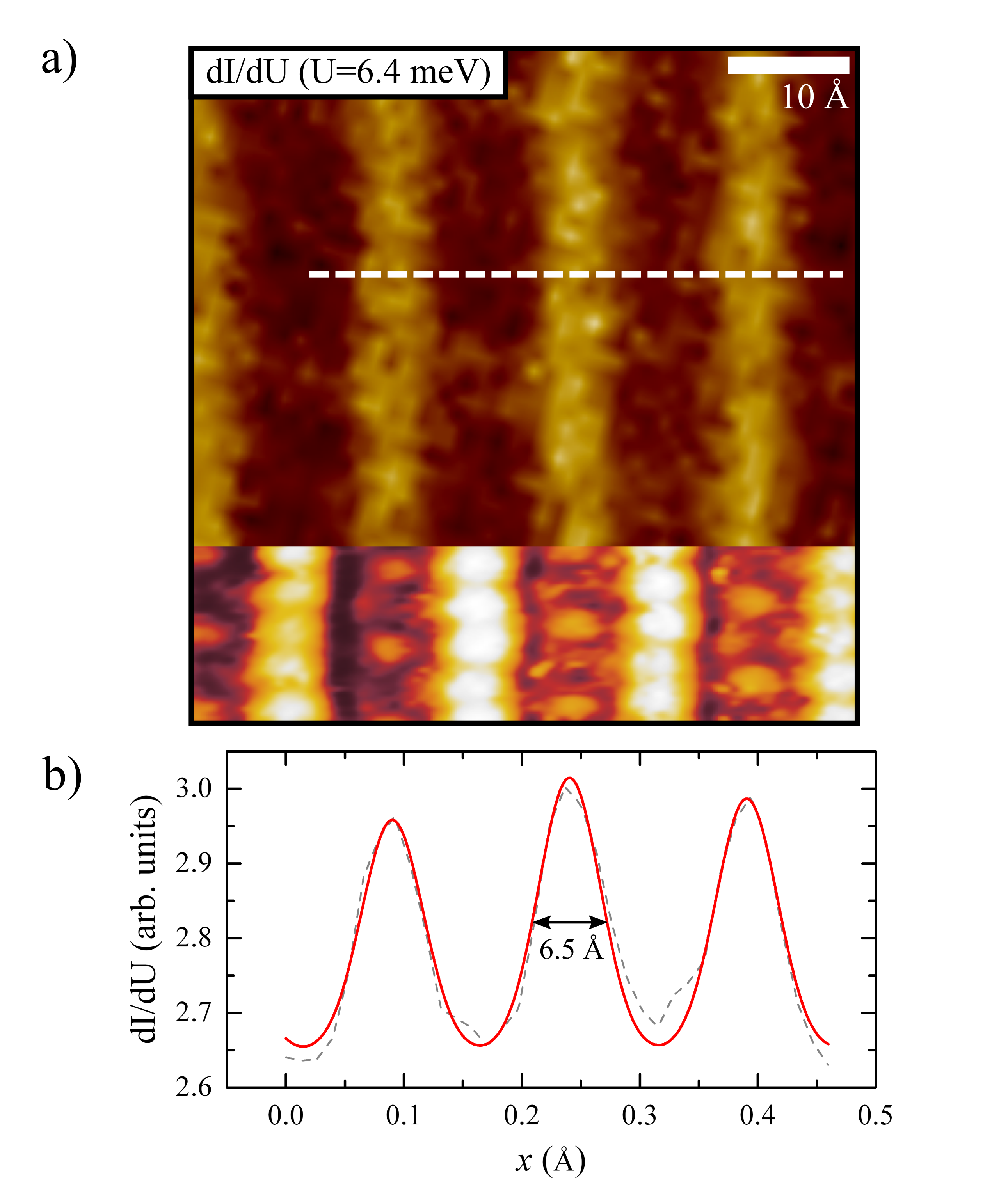}
		\caption{\label{STM} a) Combined image of tunneling microscopy and spectroscopy of Au on Si(553). 
		Bottom: topography image (U= 0.1\,V, 50\,pA) with Si edge (bright) and period doubled Au chains, very similar 
		to Si(775)-Au (see fig.\,\ref{FIG1}). Top: dI/dU map of the same sample area recorded with the Lock-In 
		technique ($U_{mod} = 10$\,mV) displayed close to $E_F$ (U$_{dc} = +6.4$\,meV) .
The density of states (DOS) is significantly enhanced at the Au chain position. 
  b) Line scan through the dI/dU map from above normal to the steps 
		indicated by the white dashed line, fit is given by the red curve.}  
\end{center}
\end{figure}

Our plasmon analysis suggests an internal modulation of the relevant electron density within each terrace. 
Although this electronic modulation may not be exactly the same in the collectively excited state, 
such a modulation is indeed seen in tunneling spectroscopy (STS) close to the Fermi level, using the Si(553)-Au system as a test
system.  This corroborates our suggestions from above. 
The combination of STM and STS (fig.\,\ref{STM}) not only shows the modulation, it demonstrates that the highest density of states (DOS)
is indeed located at the gold chains. The FWHM of this modulation (see fig.\,\ref{STM}b)) is $6.5\pm0.5$\,\AA, i.e. it is close to the geometric 
width of a double gold chain on a Si(111) terrace. Although the amplitude of this modulation is only around 10\,\%, 
it clearly demonstrates that also for the plasmon excitation this electronic density modulation must be relevant. If the plasmon of the lowest 
subband is also confined to the gold double-chain, a comparable width is expected. 
Indeed, using eq.\,\ref{eq2} and setting $a_0$ to the separation of atomic rows on the Si(111) terraces of 3.32\,\AA,
best fits of the measured dispersion curves are obtained 
with values of $a=7.5$\,\AA\ for Si(553)-Au, 10.2\,\AA\ for Si(775)-Au and 5.9\,\AA\ for Si(557)-Au. The latter value
was obtained by neglecting coupling between wires in the fit. These values suggest that confinement of quasi-1D plasmons  
is related, but not simply determined by the geometric width of the electronic ground state, and that both the 
structural motif and the terrace width play a crucial role.  
		
%%%%%%%%%%%%%%% 

%************************************************************************
%\section{Summary and Conclusion}

Summarizing, we investigated plasmonic excitations in the narrowest quasi-1D systems experimentally possible, i.e. in  
arrays of atomic metallic wires formed by single and double Au chains in vicinal Si(111) surface at various step 
densities. While only a 1D dispersing plasmon mode along the wires was found, the slope of the dispersion explicitly 
depends on the charge distribution within each mini-terrace and on distance between wires, even for identical 1D ratios
$n/m^\star$, thus superimposing 2D properties onto the 1D dispersion. While these findings require extensions of 1D plasmon theory, 
this dimensional crossover leads to new possiblities for tuning 1D plasmon dispersions.

%both Si(553)-Au and Si(775)-Au are 1d metallic systems. Having the same structural buildup and similar 
%electron densities with the same effective masses, only the terrace width is the major difference. Dispersion 
%relations have been derived from angle resolved electron energy loss spectroscopy. They show a strong difference 
%in inclination that is attributed to the lateral distribution of electron density. This distribution was confirmed 
%by STS measurements. Comparing to Si(557)-Au its modulation also depends on the atomic structure, i.e. single 
%or double chain.

%-------------------------------------------------------------------------------------------------------

\vspace{1ex}
{\bf Acknowledgement} Financial support by the Deutsche Forschungsgemeinschaft through FOR1700 is gratefully acknowledged.\\
%************************************************************************
% Bibliography
%\bibliography{Bib}

\begin{thebibliography}{99}
\bibitem{Schwarz14} B. Schwarz, P. Reininger, D. Ristani$\rm \acute{c}$,  H. Detz, A. M. Andrews, W. Schrenk, G. Strasser, Nat. Comm. {\bf 5}, 2013.
\bibitem{Atwater10} H.A. Atwater and A. Polman, Nat. Mat. {\bf 9}, 2015 (2010).
\bibitem{Oroczo14} Ciceron Ayala-Orozco et.al. Nano {\bf 8}, 6372 (2014).
\bibitem{Nagao10} T. Nagao, G. Han, C. V. Hoang, J.-S. Wi, A. Pucci, D. Weber, F. Neubrech, V.M. Silkin, D. Enders, O. Saito M. Rana, Sci. Technol. Adv. Mater. {\bf 11}, 054506 (2010).
\bibitem{Vattuone13} L. Vattuone, M. Smerieri, T. Langer, C. Tegenkamp, H. Pfn\"ur, V. M. Silkin, E. V. Chulkov, P. M. Echenique, M. Rocca, Phys. Rev. Lett.  {\bf 110}, 127405 (2013).
\bibitem{Sarma85} S. Das Sarma, Wu-yan Lai, Phys. Rev. B {\bf 32}, 1401 (1985).
\bibitem{Nagao07} T. Nagao, S. Y. Agumina, T. Inaoka, T. Sakurai,  D. Jeon, J. phys. Soc. Jap. {\bf 76}, 107714 (2007).
\bibitem{Rugeramigabo10} E.P. Rugeramigabo, C. Tegenkamp, H. Pfn\"ur, T. Inaoka, T. Nagao,  Phys. Rev. B {\bf 81}, 165407 (2010).
\bibitem{Krieg13} U. Krieg, C. Brand, C. Tegenkamp, H. Pfn\"ur , Journal of Physics: Condensed Matter {\bf 25}, 014013 (2013).
\bibitem{Diaconescu07} Diaconescu et.al. , Nature {\bf 448}, 57 (2007).
\bibitem{Smerieri14} M. Smerieri, L. Vattuone, L. Savio, T. Langer, C. Tegenkamp, H. Pfn\"ur, V.M. Silkin, M. Rocca, Phys. Rev. Lett. {\bf 113}, 186804 (2014).
\bibitem{Lindhard54} J. Lindhard, Kgl. Danske Videnskab. Selskab, Mat. Fys. Medd. {\bf 28}, No 8 (1954).
\bibitem{Stern67} F. Stern, Phys. Rev. Lett. {\bf18}, 546 (1967).
\bibitem{Singwi68} K. Singwi, M. Tosi, R. Land, A Sj\"olander, Phys. Rev. {\bf 176}, 589–599 (1968).
\bibitem{Inaoka07} T. Inaoka and T. Nagao, Material Transactions, {\bf 48}, 718 (2007).
\bibitem{Erwin10} S.C Erwin and F.J. Himpsel, Nat. Comm. {\bf 1}, 58 (2010).
\bibitem{Aulbach13} J. Aulbach, J. Sch\"afer, S.C. Erwin, S. Meyer, C. Loho, J. Settelein, R. Claessen, Phys. Rev. Lett. {\bf 111}, 137203 (2013).
\bibitem{Aulbach15} J. Aulbach, S. Erwin, R. Claessen,  J. Sch\"afer, in preparation.
\bibitem{Crain04} J.N. Crain, J.L. McChesney, Fan Zheng, M.C. Gallagher, P.C. Snijders, M. Bissen, C. Gundelach, S.C. Erwin, and F.J. Himpsel,  
Phys. Rev. B{\bf 69}, 125401 (2004).
\bibitem{Nagao06} T. Nagao, S. Yaginuma, T. Inaoka, T. Sakurai, Phys. Rev. Lett. {\bf 97}, 116802 (2006).
\bibitem{Krawiec10} M. Krawiec, Phys. Rev. B {\bf 81}, 115436 (2010).
\bibitem{Barke09} I. Barke, F. Zheng, T. K. R\"ugheimer, and F. J. Himpsel, Phys. Rev. Lett. {\bf 97},  226405 (2006).
\bibitem{Li90} Q. Li, S. Das Sarma, Phys. Rev. B {\bf 41}, 10268 (1990).
\bibitem{DasSarma96} S. Das Sarma and E.H. Hwang, Phys. Rev. B {\bf 54}, 1936 (1996).
\bibitem{Claus92} H. Claus,  A. B\"{u}ssensch\"{u}tt, M. Henzler,  Rev. Sci. Instrum. {\bf 63}, 2195 (1992).
\bibitem{Sauerbrey59} G. Sauerbrey, Zeitschrift f\"{u}r Phys. {\bf 155}, 206 (1959).
\bibitem{Krieg15} U. Krieg, T. Lichtenstein, C. Brand, C. Tegenkamp, H. Pfn\"ur, New J. Phys., {\bf  17}, 043062 (2015).
\bibitem{Krieg14} U. Krieg, Yu Zhang, C. Tegenkamp, H. Pfn\"ur, New J. Phys. {\bf 16}, 043007 (2014).


%These items are not cited:
%\bibitem{Wollschlaeger07} J. Wollschl\"ager and C. Tegenkamp, Phys. Rev. B {\bf 75}, 245439 (2007).
%\bibitem{Hogan15} Conor Hogan, private communication. Details will be published.
%\bibitem{Liu08} C. Liu, T. Inaoka, S. Yaginuma, T. Nakayama, M. Aono, and T. Nagao, Phys. Rev. B {\bf 77}, 205415 (2008).
%\bibitem{Liu08a} C. Liu, T. Inaoka, S. Yaginuma, T. Nakayama, M. Aono, and T. Nagao, Nanotechnology {\bf 19}, 355204 (2008).


\end{thebibliography}

\end{document}